\documentclass[showpacs,twocolumn,pre]{revtex4}
\usepackage[dvips]{epsfig}
\newcommand{\clR}{\mathcal{R}}

\newcommand{\clP}{{\cal P}}
\newcommand{\clB}{{\cal B}}

\newcommand{\clW}{{\cal W}}
\newcommand{\clL}{{\cal L}}
\newcommand{\clF}{{\cal F}}

\newcommand{\clE}{\mathcal{E}}

\newcommand{\prt}{\partial}

\newcommand{\frD}{D_{\rm fr}}
\newcommand{\fracI}{\,{}_0I}

\newcommand{\vep}{\varepsilon}
\newcommand{\be}{\begin{equation}}
\newcommand{\ee}{\end{equation}}
\newcommand{\bea}{\begin{eqnarray}}
\newcommand{\eea}{\end{eqnarray}}

\begin{document}

\title{A toy model of fractal glioma development under RF electric field treatment}
%\shorttitle{A toy model of fractal glioma development}

\author{A. Iomin}
%\shortauthor{A. Iomin}

\affiliation{Department of Physics, Technion, Haifa, 32000,
Israel}

\begin{abstract}
 A toy model for glioma treatment by a radio frequency
electric field is suggested. This low-intensity,
intermediate-frequency alternating electric field is known as the
tumor-treating-field (TTF). In the framework of this model the
efficiency of this TTF is estimated, and the interplay between the
TTF and the migration-proliferation dichotomy of cancer cells is
considered. The model is based on a modification of a comb model
for cancer cells, where the migration-proliferation dichotomy
becomes naturally apparent. Considering glioma cancer as a fractal
dielectric composite of cancer cells and normal tissue cells, a
new effective mechanism of glioma treatment is suggested in the
form of a giant enhancement of the TTF.  This leads to the
irreversible electroporation that may be an effective non-invasive
method of treating brain cancer.
\end{abstract}
\pacs{ {05.40.Fb}{Random walks and Levy flights},
 {87.50.S-}{Radiofrequency/microwave fields effects },
 {87.17.Ee}{Growth and division}}

\maketitle

\section{Introduction}
Recent progress in treating glioma using an alternating, radio
frequency electric field opens new questions about the
migration-proliferation dichotomy of cancer cells extended to RF
electric field.  It was shown in \textit{in vitro} and \textit{in
vivo} experiments that a low-intensity, intermediate-frequency
(100-300 kHz), alternating electric field, known as the
tumor-treating field (TTF), destroys cells that are undergoing
division \cite{palti1,palti2,palti3}. As explained, the modality
of the TTF treatment is physical and related to dielectrophoresis
(see \textit{e.g.}, \cite{pohl}), when the non-uniformity of the
TTF exerts a force, focusing at the narrow cytoplasmatic bridge
between two daughter cells at fission. This leads to disruption of
the cell proliferation, and, eventually, the cells are destroyed
without the quiescent cells of normal tissues being affected
\cite{palti1,palti2,palti3}.

One of the main features of malignant brain cancer is the ability
of tumor cells to invade the normal tissue away from the
multi-cell tumor core, and this motility is the most critical
feature of brain cancer, causing treatment failure \cite{re}.
Clinical investigations of glioma show that the proliferation rate
of migratory cells is essentially lower in the invasion region
than in the tumor core \cite{Giese1,Giese2}. This phenomenon is
known as the \textit{migration-proliferation dichotomy}, which is
an inherent property of glioblastoma (glioma), making it a highly
aggressive and invasive tumor. A question that may be of great
importance for glioma treatment is whether the TTF is effective
against invasive cells in the presence of the
migration--proliferation dichotomy, when switching between
migrating and proliferating phenotypes takes place. The switching
process between these two phenotypes is still not well understood,
and relevant models with different switching mechanisms are
suggested. Comprehensive reviews of these sophisticated approaches
can be found \cite{Khain,Deutsch,Chauviere,kolobov,fi2011}.
Therefore, for the medical treatment of cancers, dielectrophoresis
of live cancer cells can be a new direction towards understanding
this switching, since the polarization properties of migrating and
dividing cells differ \cite{feldman,khan}, and this great
selectivity can be manipulated. One of these experimental
possibilities is dielectric spectroscopy, which is a powerful tool
for investigating the dielectric properties of cells and the
structural parts of cells, and can provide important knowledge
about different cell structures, and resolve the different
properties of normal and malignant cells \cite{feldman,khan}.

Therefore, an essential question is how the TTF affects aggressive
migrating cells in the outer-invasive region with a low-rate of
proliferation, and what should be done to make the TTF an
effective means of manipulating the cancer in this region. To shed
light on this situation, we consider a simplified toy model, where
a mathematical formulation of the migration-proliferation
dichotomy can be performed in the framework of continuous-time
random walks (CTRW) \cite{shlesinger,klafter}. The simplest
mathematical realization of the CTRW mechanism of the
migration--proliferation dichotomy was introduced  for a comb
model \cite{iom2006}, which is a toy model, where the migration
proliferation dichotomy is naturally apparent. The collective
behavior of cells was considered and the primary focus was on the
influence of cell fission on the transport properties of cells.

\section{Cell kinetics in presence of the TTF}
In the present study, we first consider the kinetic properties of
cancer development in the presence of the TTF. These can be
described by fractional kinetics in the framework of the comb
model, where it is easier to draw an intelligible picture of
interplay between high-motility aggressive cancer cells and the
TTF in the outer-invasive region. Following toy model
consideration \cite{iom2006}, we restrict our model to one
dimensional subdiffusion in the $x$ direction. In the comb model,
this anomalous diffusion can be described by the $2D$ distribution
function $P=P(x,y,t)$, and a special behavior is that the
displacement in the $x$--direction is possible only along the
structure axis ($x$-axis at $y=0$). The Fokker-Planck equation in
some dimensionless variables reads $$ \prt_tP= \delta(y) \prt_x^2P
+d\prt_y^2P\, , $$ where $d$ is an effective diffusion
coefficient. Obviously, cell division is random in $x$ and
discontinuous, contrary to that in the tumor core. It can be
reasonably considered as a fractal set $F_{\nu}(x)$ with the
fractal dimension $\nu$, which is embedded in the $1D$ space,
$0<\nu<1$. Therefore, the effective diffusion coefficient becomes
inhomogeneous $d\rightarrow d\chi(x)$, where $\chi(x)$ is a
characteristic function of the fractal, such that $\chi(x)=1$ for
$x\in F_{\nu}(x)$ and $\chi(x)=0$ for $x\notin F_{\nu}(x)$. Now we
take into account the influence of the TTF that affects (destroys)
only the proliferation phenotype according to
Refs.~\cite{palti1,palti2,palti3}. Mathematically, this process is
expressed by diffusion in the $y$ direction with decay: $$
d\prt_y^2P(x,y,t)\Rightarrow \chi(x)[d\prt_y^2-C]P(x,y,t)\, ,$$
where coefficient $C$ defines difference between the proliferation
and the degradation rate. Taking this into account, one arrives at
a model that makes it possible to understand the influence of the
TTF on the cancer development
\be\label{f_gl3}  %
\prt_tP=\delta(y)\prt_x^2P+\chi(x)[d\prt_y^2-C]P(x,y,t)\, . \ee
First, we apply the Fourier transform to Eq. (\ref{f_gl3}) with
respect to the $x$ coordinate. To apply this transformation to the
last term in Eq. (\ref{f_gl3}), we use the auxiliary identity
$\chi(x)f(x)\equiv\prt_x\int_{-\infty}^x\chi(y)f(y)dy\equiv
-\prt_x\int_x^{\infty}\chi(y)f(y)dy$ with the boundary conditions
$P(x=\pm\infty)=0$. This integration with the characteristic
function can be carried out by means of a convolution
\cite{bi2011a,iom2011,Ren}\footnote{Note, $
\int_{-\infty}^{\infty}\chi(y)f(y)dy=\sum_{x_j\in
F_{\nu}}\int_{-\infty}^{\infty}f(y)\delta(y-x_j)dy$, where
$\sum_{x_j\in F_{\nu}}\delta(y-x_j)=\mu^{\prime}(x)\sim
|x|^{\nu-1}$ is a fractal density, such that
%on the finite interval $(-x,x)$, the integral
$\int_{-x}^xd\mu(y)\sim |x|^{\nu}$ corresponds to the fractal
volume. Therefore, due to Theorem $3.1$ in Ref.~\cite{Ren} we have
$\int_0^xf(y)d\mu(y)\simeq
\frac{1}{\Gamma(\nu)}\int_0^x(x-y)^{\nu-1}f(y)dy$.
%, which is defined for the finite fractal volume $\mu(x)$.
}. %
To this end we will use the terminology and useful notations of
fractional integration and differentiation
\cite{klafter,oldham,podlubny,WBG,SKM,SokKlafBlum}. Fractional
integration of the order of $\nu$ is defined by the operator (see
also Appendix)
\be\label{f_gl4}   %
{}_{-\infty}I_x^{\nu}f(x)=
\frac{1}{\Gamma(\nu)}\int_{-\infty}^xf(y)(x-y)^{\nu-1}dy\, , \ee %
where $0<\nu<1$ and  $\Gamma(\nu)$ is the Gamma function. The Weyl
fractional derivative is
$$\clW_x^{1-\nu}=\prt_x[{}_{-\infty}I_x^{\nu}f(x)]
={}_{-\infty}I_x^{\nu}[\prt_xf(x)]\, .$$ Using this notation, one
introduces the coarse-graining integration with the characteristic
function in the form of the Weyl fractional derivative
\cite{bi2011a}
\be\label{f_gl5}  %
\chi(x)P(x,y,t)\Rightarrow\clW_x^{1-\nu}P(x,y,t)\,  .  \ee   %
The Fourier transform of the Weyl derivative yields
$$\hat{\clF}_x\clW_x^{1-\nu}P(x,y,t)=(ik)^{1-\nu}\hat{P}(k,y,t)\,
,$$ where $\hat{\clF}_x[P(x,y,t)]=\hat{P}(k,y,t)$. Presenting this
Fourier transform in the symmetrical form, one obtains
\be\label{f_gl6}  %
\prt_t\hat{P}=-k^2\delta(y)\hat{P}+
|k|^{1-\nu}[d\prt_y^2\hat{P}-C\hat{P}]\, . \ee   %
The last term in the r.h.s. of Eq. (\ref{f_gl6}) is eliminated by
substituting $\hat{P}(k,y,t)=e^{-C|k|^{1-\mu}}F(k,y,t)$. Then one
carries out the Laplace transform in the time domain
$\hat{\clL}[F(k,y,t)]=\tilde{F}(k,y,s)$. Looking for the solution
of the Laplace image in the form
\be\label{f_gl7}  %
\tilde{F}(k,y,s)=\exp[-|y|\sqrt{|k|^{\nu-1}s/d}]f(k,s)\, ,  \ee  %
one arrives at the intermediate expression in the form of the
Laplace and Fourier inversions
\be\label{f_gl8}  %
P(x,y,t)=\hat{\clF}_k^{-1}\left\{e^{-C|k|^{1-\nu}t}\hat{\clL}^{-1}
\left[ \frac{2e^{-|y|\sqrt{s|k|^{\nu-1}/d}}}{2\sqrt{sd
|k|^{1-\nu}}+ k^2}\right]\right\}\, . \ee %
One has to recognize that the $y$ axis is the auxiliary
coordinate, which determines the cell proliferating process (cell
fission). Therefore to find the complete distribution of cancer
cells in the $x$ axis, integration over $y$ is performed:
$\overline{P}(x,t)=\int_{-\infty}^{\infty}P(x,y,t)dy$. Both the
integration over $y$ and the inverse Laplace transform are carried
out exactly. This, eventually, yields a solution in the
convolution form
\bea %
&\overline{P}(x,t)=\frac{1}{(Ct)^{\frac{1}{1-\nu}}}
\int_{-\infty}^{\infty}\clR\Big[\frac{x-x'}{(Ct)^{\frac{1}{1-\nu}}}\Big]
\clP\Big(\frac{x'}{t^{\frac{1}{3+\nu}}}\Big)dx'
\label{f_gl9a} \\
&=\frac{1}{2\pi}\int_{-\infty}^{\infty}e^{-ikx}
e^{-C|k|^{1-\nu}t}\clE_{\frac{1}{2}}\Big(-\frac{1}{2}\sqrt{|k|^{3+\nu}t/d}\Big)dk
\, . \label{f_gl9b} \eea  %
Here $$\clE_{\alpha}(-z)=\frac{1}{2\pi
i}\int_{\gamma}\frac{u^{\alpha-1}e^udu}{u^{\alpha}+z}$$ is the
Mittag-Leffler function defined by the inverse Laplace transform
with a corresponding deformation of the contour of the integration
\cite{batmen}, while $\clP(z)$ is a solution for the untreated
glioma with the scaling variable $z=x/t^{\frac{1}{3+\nu}}$ and
$\clR(z)$ is the kernel of the TTF treatment. For the small
argument, which corresponds (for a short time) to a long-scale
tail of the distribution, the Mittag-Leffler function decays
exponentially $\exp\Big(-K_{\nu}\sqrt{|k|^{3+\nu}t}\Big)$
\cite{klafter,batmen} with the generalized transport coefficient
$K_{\nu}=[2\Gamma(3/2)\sqrt{d}]^{-1}$. This yields the solution
for the untreated cancer in the form of the Fox functions
\cite{klafter}. Its large argument asymptotic solution yields the
power law decay $ \clP(z)\sim \sqrt{t/|x|^{5+\nu}}$ that is an
indication of the non-localized diffusive cancer in the
outer-invasive region\footnote{This is a mathematical expression
of a statement that the surgical resection is ineffective since
the cancer cells have already invaded into the surrounding brain
tissue. This leads to recurrence of tumor and prognosis for
patients suffering from malignant gliomas is very poor
\cite{re,Giese1,Giese2}. }.

Now we turn to the non-invasive treatment in Eqs. (\ref{f_gl9a})
and (\ref{f_gl9b}), described by the kernel
$$\frac{1}{(Ct)^{\frac{1}{1-\nu}}}
\clR\Big[\frac{x}{(Ct)^{\frac{1}{1-\nu}}}\Big] =
\hat{\clF}_k^{-1}\Big[e^{-C|k|^{1-\nu}t}\Big]\, ,$$ which is the
Fox function. Therefore, the resulting solution in Eqs.
(\ref{f_gl9a}) and (\ref{f_gl9b}) is the convolution of two Fox
functions, which means that the TTF is a fairly inefficient
treatment in the outer-invasive region. Our main aim here is to
understand the efficiency of the TTF. As seen from Eqs.
(\ref{f_gl9a}) and (\ref{f_gl9b}), the crucial point is the
fractal dimension $\nu<1$. For $\nu=1$ the solution is
$$\overline{P}(x,t)=e^{-Ct}\clP(x^2/t^{\frac{1}{2}})\, ,$$
where $\clP(z)$ is the stretched exponent. This means that the TTF
is most efficient in the multi-cell tumor core and in the close
vicinity of the core, where the proliferating cell distribution is
continuous with the fractal dimension $\nu=1$. Therefore, to
increase the efficiency of the TTF in the outer-invasive region,
the TTF must act on both the migrating and the proliferating
cells.

\section{Electroporation}
In what follows we suggest a possible mechanism of such a
treatment of the diffusive cancer using the TTF in the
outer-invasive region. The idea is based on the differences
between the dielectric properties of normal tissue cells and
cancer cells that can lead to the local enhancement of the
electric field and to electroporation of the cancer cells.
Therefore, we consider the frequency-dependent permittivities of
migrating cancer cells $\vep_m$ and the normal tissue cells
$\vep_n$. Under certain frequency, the condition $\vep_m<\vep_n$
can be fulfilled. This was found for dielectric properties of
normal, and malignant lymphocytes \cite{feldman}. Therefore, the
outer-invasive region is a fractal composite of two dielectrics
with permittivities $\vep_n$ and $\vep_m$, where
$\vep_n=\vep_n(\omega)$ corresponds to the host dielectric, while
$\vep_m=\vep_m(\omega)$ corresponds to the fractal cancer
inclusion. We do not consider proliferating cancer cells, since
the FFT destroys them effectively. Note also the wavelength of the
TTF is much larger than any brain inhomogeneities, and therefore
the quasi-electrostatic consideration is valid. The host volume of
normal tissue cells is $3D$, while the volume of cancer cells is
fractal $V\sim r^{\frD}$ with fractal dimension $\frD<3$ due to
their low concentration, which is much less than their
concentration in a solid tumor. Therefore, the outer-invasive
region can be considered as a random fractal cancer with an
averaged fractal density $\rho(r)\sim r^{\frD-3}$, as
\textit{e.g.}, for low-grade astrocytomas. Note, in
electrostatics, $\rho(r)$ is considered as a fractal quenched
disorder distribution.

Electrostatics in the frequency domain (when
$\mathbf{E}=\mathbf{E}(\omega,\mathbf{r})$), in the random fractal
dielectric composite is described by the Maxwell equation
\be\label{el_gl1}  %
\nabla\cdot\Big[\bf{E}\vep(\mathbf{r})\Big]=0\, , \ee %
where the inhomogeneous permittivity is a function of the
frequency of external TTF and coordinates:
$\vep(\mathbf{r})=\vep(\mathbf{r},\omega)$. The boundary
conditions are determined by the external TTF. We work with
dimensionless coordinates, which are scaled by the cell size. It
is convenient to split the electric field inside the brain into
two components
$\mathbf{E}(\mathbf{r})=\mathbf{\bar{E}}(\mathbf{r})+\mathbf{E}_1(\mathbf{r})$,
where $\mathbf{\bar{E}}$ is the homogeneous electric field with
the condition $\nabla\cdot\mathbf{\bar{E}}=0$, while
$\mathbf{E}_1$ is the electric field due to fractal cancer
inhomogeneities. We take into account that the fractal mass
$M(r)\sim r^{\frD}$ results from averaging over all the possible
realizations of the random fractal. The averaged density
distribution $\rho(r)$ is isotropic and depends on the radius only
(in complete agreement with the probabilistic sense of random
fractals \cite{berrypercival,benavraam}). This also supposes that
the dielectric properties of the random fractal composite are
isotropic, such that the space-dependent averaged polarization is
a function of the radius only: $\vep(\mathbf{r})=\vep(r)$.
Therefore, it is reasonable to suppose that the response to the
TTF due to the fractal inhomogeneities changes as a function of
the radius ${\bf E}_1(\omega,\mathbf{r})=\mathbf{E}_1(\omega,r)$,
as well. Thus the divergence in the Maxwell equation for ${\bf
E}_1$ takes into account only the $r$ component of the electric
field. This yields $\nabla\cdot{\bf E}_1= \nabla_rE_1(r)$, where
$E_1(r)$ is the $r$ component of the electric field (the $r$ index
is omitted). The inhomogeneous permittivity of the composite we
define by means of the fractal characteristic function
$\chi(r)=1$, if $r$ belongs to cancer cells, and $\chi(r)=0$, if
$r$ belongs to the normal tissues. Therefore
$\vep(r)=\vep_m\chi(r)+[1-\chi(r)]\vep_n\equiv
\vep_n[\xi\chi(r)+1]$, where we introduced  dimensionless
parameter $\xi=(\vep_m-\vep_n)/\vep_n$. Substituting this in the
Maxwell equation (\ref{el_gl1}), one obtains
\be\label{el_gl2}  %
[\xi\chi(r)+1]\nabla_rE_1+E_1\chi^{\prime}(r)=-\bar{E}_r\chi^{\prime}(r)\,
.   \ee  %
where $\chi^{\prime}(r)\equiv \frac{d}{dr}\chi(r)$, and we take
into account that $\nabla_r=\frac{1}{r^2}\frac{d}{dr}r^2$. We
introduce a new function $\clB(r)=r^2E_1(r)$, and then the
integration of Eq. (\ref{el_gl2}) with the elementary volume $
dV=4\pi r^2dr$ reads
\bea\label{el_gl3}  %
\xi\int_0^r\chi(r)\clB^{\prime}dr&+&\clB
+\xi\int_0^r\clB\chi^{\prime}(r)dr= \nonumber  \\
&-& \xi \int_0^r\bar{E}_r(r)r^2\chi^{\prime}(r)dr\, ,\eea  %
where $\clB^{\prime}(r)\equiv\frac{d}{dr}\clB(r)$.

Now we in a position to use fractional calculus for integration
with the fractal characteristic function. By complete analogy with
Eqs. (\ref{f_gl4}) and (\ref{f_gl5}), this integration corresponds
to the convolution integral with the averaged density $\rho(r)$
that yields \cite{bi2011a}
\begin{eqnarray*}
\int_0^r\chi(r)f(r)dr &\Rightarrow& \frac{1}{\Gamma(\frD-2)}
\int_0^r(r-r')^{\frD-3}f(r')dr' \\
&\equiv& \fracI_r^{\frD-2}f(r)\, ,
\end{eqnarray*}
where $f(r)$ is arbitrary, in particular, $f(r)=\clB^{\prime}(r)$.
The situation with polarization terms is more complicated, and one
ought to take into account the electrostatic problem of
polarization of dielectric balls by an external electric field
\cite{pohl}. All details of the inferring can be found in
Refs.~\cite{bi2011a,bi2011b}, and the finite result reads
$$
\int_0^r\chi'(r)\clB(r)dr\Rightarrow p\fracI_r^{\frD-2}\clB(r)\,
,$$ where $p=\frac{\xi^2}{3+\xi}$ is a polarization parameter that
is well known in the literature \cite{LLVIII,bohren,pohl}. Using
this expression and taking into account that in homogeneous media,
when cancer is absent, $\bar{E}=E_0/\vep_n$, where
$E_0=E_0(\omega)$ is the amplitude of the TTF in the frequency
domain, one easily calculates the term in the r.h.s. of Eq.
(\ref{el_gl3}):
$$\frac{pE_0\xi}{\vep_n}\fracI^{\frD-2}r^2=
\frac{2pE_0\xi}{\vep_n\Gamma(\frD+1)}r^{\frD}\, .$$ Now Eq.
(\ref{el_gl3}) can be rewritten in the form convenient for solving
\be\label{el_gl4}  %
\xi\fracI_r^{\frD-2}\clB^{\prime}+\clB+p\fracI_r^{\frD-2}\clB=
-\frac{2pE_0\xi}{\vep_n\Gamma(\frD+1)}r^{\frD}\, .  \ee  %
We consider a case with negative $\xi$ and  $|\xi|\ll 1$, which is
the most important for the electric responce to the TTF. In this
case one neglects the $p\fracI_r^{\frD-2}\clB$ term. The Laplace
transform method yields the solution in the form of the
two-parameter Mittag-Leffler function \cite{batmen}
$$\clE_{\alpha,\beta}\Big(\frac{r^{\alpha}}{|\xi|}\Big)=
\frac{r^{1-\beta}}{2\pi i} \int_{\gamma}
\frac{u^{\alpha-\beta}e^{ur}du}{u^{\alpha}-1/|\xi|}\, , $$ where,
in our case, $\alpha=3-\frD$, $\beta=4$. For small $|\xi|$, the
large argument asymptotics of the Mittag-Leffler function is
exponential \cite{batmen}: $\clE_{\alpha,\beta}(z)\sim
\frac{1}{\alpha}z^{-3}e^z$, where $z= \frac{r^{\alpha}}{|\xi|}$.
This, finally, yields the solution for the electric field
\be\label{el_gl5}  %
E_1(r)\sim \frac{E_0}{\vep_n}\cdot
\frac{|\xi|^{\frac{3+\alpha}{\alpha}}}{r^2}
\exp\Big(\frac{r}{|\xi|^{\frac{1}{\alpha}}}\Big)\, ,
~~\alpha=3-\frD\, . \ee  %
Therefore, the respond electric field can be large enough to break
the cell membrane. For example, for $|\xi|\sim 0.2$, $\frD=2.5$,
and $\vep_n~\sim 10^2$, the electric field response is $10^4\div
10^5$ V/cm, which exerts the irreversible electroporation
\cite{Rubinsky} due to the external TTF with amplitude $E_0\sim
1$V/cm. This can be a mechanism for ablation of cancer cells,
which effectively acts on migratory cancer cells. An important
condition for this realization is $\vep_m(\omega)<\vep_n(\omega)$.
These permittivities were observed in time domain dielectric
spectroscopy in  experimental studies of the static and dynamic
dielectric properties of normal, transformed, and malignant B- and
T-lymphocytes \cite{feldman}. Such experimental studies of the
glioma cells can not be overestimated.

\section{Conclusion}
We presented two models of treating glioma  by alternating, radio
frequency electric field, which is the tumor-treating field (TTF)
in the presence of the migration-proli\-fe\-ra\-tion dichotomy of
cancer cells. The first model considers the treating tumor
development on a comb model. In the framework of this toy model of
the migration-proli\-fe\-ra\-tion dichotomy, based on the
fractional cell transport, it was possible to estimate the
effectiveness of the TTF treatment in the outer-invasive region of
the tumor development. We also show that while the TTF is highly
effective in the multi-cell tumor core, its action is ineffective
in the presence of the migration-proliferation dichotomy. The key
reason is the fractal structure of the glioma cancer in the
outer-invasive region. Therefore, another possible mechanism of
the diffusive cancer treatment by the low-intensity alternating
electric field has been suggested. The idea is based on the
difference between the dielectric properties of normal tissue
cells and cancer cells \cite{feldman,khan}. Since the cell
permittivities are functions of the TTF frequency, it is supposed
that there are conditions where the permittivity of normal cells
is larger than that of cancer cells \cite{feldman}. Therefore,
considering the tumor invasion through normal tissue as a fractal
dielectric composite in the presence of a high-frequency electric
field, the electrostatic Maxwell equation in a fractal medium was
considered. An analytical solution was obtained in the framework
of fractional calculus, and an essential enhancement of the
electric field was obtained. The result depends essentially on the
cancer fractal dimension $\frD$ and the difference between the
permittivities $|\xi|<1$. In view of the toy model (\ref{f_gl3}),
this treatment leads to the solution $\overline{P}(x,t)\sim
e^{-Ct}\clP(x,t)$, where $C$ is electroparation, which is an
effective treatment in the invasive region.

A key quantity of the cancer treatment is localization of the
electroporation field $E_1(r)$ inside the cancer. There is a
straightforward analogy with nanoplasmonics (see \textit{e.g.},
\cite{Shalaev,Stockman}), where the electric field enhancement is
due to a so-called surface-plasmon resonance for a
metal-dielectric composite, and localized surface plasmon
oscillations are charge density oscillations confined to the
conducting fractal nanostructure. The essential difference and
similarity should be admitted: this biological, cell enhancement
of the electric field is not resonant, but geometrical due to the
fractal cancer structure \cite{bi2011b}. In this connection,
\textit{in vitro} experiments can be important for further
understanding the interplay between the TTF and the
migration-proliferation dichotomy

\section{Appendix}
Fractional integration of the order of $\alpha$ is defined by the
operator
$${}_aI_x^{\alpha}f(x)=
\frac{1}{\Gamma(\alpha)}\int_a^xf(y)(x-y)^{\alpha-1}dy\, , $$
where $\alpha>0,~x>a$ and  $\Gamma(z)$ is the Gamma function. The
fractional derivative is the inverse operator to
${}_aI_x^{\alpha}$ as $ {}_aD_x^{\alpha}f(x)={}_aI_x^{-\alpha}$
and ${}_aI_x^{\alpha}={}_aD_x^{-\alpha}$. Its explicit form is
$${}_aD_x^{-\alpha}=
\frac{1}{\Gamma(-\alpha)}\int_a^xf(y)(x-y)^{-1-\alpha}dy\, .$$ For
arbitrary $\alpha>0$ this integral diverges, and as a result of a
regularization procedure, there are two alternative definitions of
${}_aD_x^{-\alpha}$. For an integer $n$ defined as $n-1<\alpha<n$,
one obtains the Riemann-Liouville fractional derivative of the
form $${}_aD_{RL}^{\alpha}f(x)=(d^n/x^n){}_aI_x^{n-\alpha}f(x)\,
,$$ and fractional derivative in the Caputo form
$${}_aD_{C}^{\alpha}f(x)= {}_aI_x^{n-\alpha}f^{(n)}(x)\, . $$ When
$a=-\infty$, the resulting Weyl derivative is
$$\clW^{\alpha}\equiv{}_{-\infty}D_{W}^{\alpha}=
{}_{-\infty}D_{RL}^{\alpha}= {}_{-\infty}D_{C}^{\alpha}\, .$$  One
also has ${}_{-\infty}D_{W}^{\alpha}e^x=e^x$ This property is
convenient for the Fourier transform
$\hat{\clF}\left[\clW^{\alpha}f(x)\right]=(ik)^{\alpha}\hat{f}(k)$,
where $\hat{\clF}[f(x)]=\hat{f}(k)$.

\end{document}